\documentclass[a4paper,12pt]{article}
\pdfoutput=1
\usepackage{jheppub}
\usepackage{datetime}
\usepackage{float}

\title{Kerr-Newman Black Holes with String Corrections}

\author{Anthony M. Charles and}
\author{Finn Larsen}

\affiliation{Department of Physics and Michigan Center for Theoretical Physics, \\
University of Michigan, 450 Church Street, Ann Arbor, MI 48109-1020, USA}

\emailAdd{amchar@umich.edu}
\emailAdd{larsenf@umich.edu}

\abstract{We study $\mathcal{N}=2$ supergravity with higher-derivative corrections that preserve the $\mathcal{N}=2$ supersymmetry and show that Kerr-Newman black holes are solutions to these theories. Modifications of the black hole entropy due to the higher derivatives are universal and apply even in the BPS and Schwarzschild limits. Our solutions and their entropy are greatly simplified by supersymmetry of the theory even though the black holes generally do not preserve any of the supersymmetry.}

\begin{document}
\maketitle
\flushbottom

%%%%%%%%%%%%%%%%%%%%%%%%%%%%%%%%%%%%%%%%%%%%%%%%%%%%%%%%%%%%%%%%%%

\section{Introduction and Summary}
\label{sec:intro}

%%%%%%%%%%%%%%%%%%%%%%%%%%%%%%%%%%%%%%%%%%%%%%%%%%%%%%%%%%%%%%%%%%

Most precision studies of black holes in string theory are carried out near the BPS limit where supersymmetry guarantees control. It is thought that various corrections become unwieldy far from this limit. Curiously, most discussions of the black hole information paradox are carried out in the opposite limit of Schwarzschild black holes since their geometries are the simplest. It is thought that this is sufficient to gain universal insights. Few details on the implied interpolation between the BPS and Schwarzschild limits are known. 

In this paper we construct families of black holes that interpolate between these limits while taking certain string corrections into account. We find that the string corrections are surprisingly manageable. The simplifications we report are due to supersymmetry of the theories we consider. Importantly, they persist even though the black holes we construct generally do not preserve any of the supersymmetry.

A convenient starting point for connection with studies that are not motivated by string theory is the 4D Einstein-Maxwell theory
\begin{equation}
	{\cal L}_{\rm EM} =  -{1\over 16\pi G_N} \left(R  + {1\over 4}  F_{\mu\nu}F^{\mu\nu}\right) ~.
\end{equation}
We primarily consider the standard Kerr-Newman family of solutions that includes BPS black holes and Schwarzschild black holes as special cases. 

A simple way to add higher-derivative terms to this theory is to consider the Gauss-Bonnet density
\begin{equation}
\label{eqn:euler}
{\cal L}_{\rm GB} = \alpha E_4 = \alpha \left( R_{\mu\nu\rho\sigma}R^{\mu\nu\rho\sigma} - 4 R_{\mu\nu}R^{\mu\nu} + R^2\right)~.
\end{equation}
This term is topological so the equations of motion are unchanged and therefore solutions remain the same. Black holes nevertheless have a different entropy in the modified theory because the Wald entropy formula depends on the action~\cite{Wald:1993nt,Iyer:1994ys,Jacobson:1993vj}. 

Generally other linear combinations of the curvature invariants are much more complicated. The Weyl invariant
\begin{equation}
\label{eqn:weyl}
{\cal L}_{\rm Weyl} = \gamma W_{\mu\nu\rho\sigma}W^{\mu\nu\rho\sigma}= \gamma \left( R_{\mu\nu\rho\sigma}R^{\mu\nu\rho\sigma} - 2 R_{\mu\nu}R^{\mu\nu} + \frac{1}{3} R^2\right)~,
\end{equation}
introduces the Bach tensor into the equations of motion which are then difficult to analyze. In our work we are inspired by string theory and consider theories with ${\cal N}=2$ supersymmetry such as the supersymmetric completion of (\ref{eqn:weyl}) that takes the schematic form (made precise in equation (\ref{eq:susyweyl_precise}) below):
\begin{equation}
\label{eqn:susyweyl}
{\cal L}_{\rm {\cal N}=2~Weyl} = \gamma_1 W^2 + \gamma_2 F^4 + \gamma_3 W F^2 + \ldots~,
\end{equation}
with various contractions of the tensors. In this case the equations of motion are even more complicated and it is not clear from the outset that it is realistic to solve them. We find that, surprisingly, any solution to Einstein-Maxwell theory automatically solves the full theory with ${\cal N}=2$ supersymmetry. This will allow us to study generic non-supersymmetric solutions in the presence of higher-derivative corrections. 

The higher-derivative corrections modify the Wald entropy of Kerr-Newman black holes. It turns out that the combined contribution from all the terms in the supersymmetrized Weyl invariant (\ref{eqn:susyweyl}) is precisely the same as the modification due to the Gauss-Bonnet density (\ref{eqn:euler}) alone. In particular, the contribution from higher-derivative terms is topological. It is therefore independent of black hole parameters and can be extrapolated arbitrarily far from the BPS limit with no change. 

The supersymmetrized Weyl invariant (\ref{eqn:susyweyl}) commonly appears in low energy effective actions. For example, it arises when massive string modes are integrated out.  (see \emph{e.g.}~\cite{Gross:1986mw,Bergshoeff:1989de,Castro:2008ne,Castro:2007hc,Castro:2007ci}.) The terms we consider are string corrections in this sense. Our result indicates that string corrections are milder than previousely suspected. 

Massless modes running in virtual loops offer a related quantum mechanism that gives higher-derivative terms at low energy. In previous work~\cite{Charles:2015eha} we studied the logarithmic corrections to Kerr-Newman entropy due to such effects. In general these logarithmic corrections are very complicated but upon embedding of the Kerr-Newman 
black hole into a theory with ${\cal N}\geq 2$ supersymmetry they greatly simplify and become independent of the black hole parameters. 

The two classes of corrections we have considered both show that black hole entropy depends greatly on the setting. In an environment with ${\cal N}\geq 2$ supersymmetry there are considerable simplifications even for black holes that do not themselves preserve any supersymmetry. Indeed, several of the corrections to the entropy that have been analyzed precisely in the BPS limit do not depend on black hole parameters at all and so apply far off extremality. This result raises hopes that the entropy of non-supersymmetric black holes can be understood precisely in a microscopic theory. 

This paper is organized as follows.  In section~\ref{sec:sugra} we present a simplified summary of off-shell $\mathcal{N}=2$ supergravity. (More details are given in the appendix.) In section~\ref{sec:minsugra} we study minimal supergravity with higher-derivative corrections in the form of a supersymmetrized Weyl invariant and derive the full equations of motion for the theory.  In section~\ref{sec:em} we embed arbitrary Einstein-Maxwell solutions into our minimal supergravity theory and show that all fields are unchanged, even for solutions that do not preserve supersymmetry.  In section~\ref{sec:properties} we study properties of black holes in this embedding and find that the correction to the black hole entropy is topological and independent of black hole parameters.  Finally, in section~\ref{sec:discussion} we discuss our results and potential implications for microscopic models of Kerr-Newman black holes.

%%%%%%%%%%%%%%%%%%%%%%%%%%%%%%%%%%%%%%%%%%%%%%%%%%%%%%%%%%%%%%%%%%

\section{Higher-Derivative $\mathcal{N}=2$ Supergravity}
\label{sec:sugra}

%%%%%%%%%%%%%%%%%%%%%%%%%%%%%%%%%%%%%%%%%%%%%%%%%%%%%%%%%%%%%%%%%%
%
%Our ultimate goal is to interpret solutions to Einstein-Maxwell theory (and, in particular, Kerr-Newman black holes) as solutions to $\mathcal{N} = 2$ supergravity in 4D.  We studied such an embedding previously~\cite{Charles:2015eha|, \cite,Gibbons:1982fy,Mohaupt:2000gc}, but now want to generalize and include higher-derivative terms.  

The details of 4D off-shell $\mathcal{N}=2$ supergravity with higher-derivative interactions have been studied exhaustively~\cite{Behrndt:1998eq,LopesCardoso:1998tkj,Mohaupt:2000mj,LopesCardoso:2000qm,LopesCardoso:2000fp}. We review some technical details in the appendix.  In this section we present a more elementary and accessible discussion of $\mathcal{N}=2$ supergravity with higher-derivative corrections. 

%%%%%%%%%%%%%%%%%%%%%%%%%%%%%%%%%%%%%%%%%%%%%%%%%%%%%%%%%%%%%%%%%%

\subsection{Field Content}
\label{subsec:sugra_whatwedo}

%%%%%%%%%%%%%%%%%%%%%%%%%%%%%%%%%%%%%%%%%%%%%%%%%%%%%%%%%%%%%%%%%%

We focus on the bosonic fields in $\mathcal{N} = 2$ supergravity. The physical $\mathcal{N} = 2$ gravity multiplet contains the metric $g_{\mu\nu}$ and a $U(1)$ graviphoton field. We further couple this theory to $n_V$ physical $\mathcal{N} = 2$ vector multiplets, each comprising a $U(1)$ gauge field and a complex scalar. The version of the off-shell formalism we employ realizes this coupling by introducing $n_V + 1$ vectors $W_\mu^I$ and $n_V + 1$ complex scalars $X^I$, where $I = 0,\ldots, n_V$. One of the complex scalars can be removed by symmetries and does not correspond to physical degrees of freedom.  Without loss of generality, we can choose the auxiliary scalar to be $X^0$, and we will index the physical vector multiplets by $a = 1,\ldots,n_V$.  The remaining gauge field $W_\mu^0$ gets combined with the metric to form the $\mathcal{N}=2$ gravity multiplet on-shell.

The complete formalism based on realization of superconformal symmetry contains many other auxiliary fields that must be carefully considered.  However, for our purposes we can consistently set most of these fields to zero at the level of the action. The only ones we must retain are a $U(1)_R$ vector field $A_\mu$, an anti-self-dual antisymmetric tensor $T^-_{\mu\nu}$ and a scalar $D$ that all belong to an off-shell $\mathcal{N}=2$ Weyl multiplet  with the metric.

We summarize this discussion with a list of fields, from both the off-shell and the on-shell perspectives, in table~\ref{table:fieldcontent}.

\bgroup
\def\arraystretch{1.2}
\begin{table}[H]
\centering
\begin{tabular}{|cl|}\hline
	\multicolumn{2}{|c|}{\textbf{Off-Shell Field Content}} \\ \hline
	Weyl multiplet: & $g_{\mu\nu}~,~A_\mu~,~T^-_{\mu\nu}~,~D~$  \\
	Vector multiplets: & $W_\mu^I~,~X^I~$ \\
	\multicolumn{2}{|c|}{$(I = 0,\ldots,n_V)$} \\ \hline
\end{tabular}\quad\quad
\begin{tabular}{|cl|}\hline
	\multicolumn{2}{|c|}{\textbf{Physical Field Content}} \\ \hline
	Gravity multiplet: & $g_{\mu\nu}~,~W_\mu^0~$  \\
	Vector multiplets: & $W_\mu^a~,~X^a~$ \\
	\multicolumn{2}{|c|}{$(a = 1,\ldots,n_V)$} \\ \hline
	\multicolumn{2}{c}{} \\ \hline
	\multicolumn{2}{|c|}{\textbf{Auxiliary Fields}} \\ \hline
	\multicolumn{2}{|c|}{$X^0~,~A_\mu~,~T^-_{\mu\nu}~,~D$} \\ \hline
\end{tabular}
\caption{Summary of the field content in the $\mathcal{N}=2$ supergravity theory.  The $n_V+1$ off-shell vector multiplets are indexed by $I$, while the $n_V$ physical vector multiplets are indexed by $a$.}
\label{table:fieldcontent}
\end{table}
\egroup

%%%%%%%%%%%%%%%%%%%%%%%%%%%%%%%%%%%%%%%%%%%%%%%%%%%%%%%%%%%%%%%%%%

\subsection{Definitions and Notation}
\label{subsec:sugra_definitions}

%%%%%%%%%%%%%%%%%%%%%%%%%%%%%%%%%%%%%%%%%%%%%%%%%%%%%%%%%%%%%%%%%%

We will denote the field strengths of the $U(1)_R$ gauge field $A_\mu$ and the $n_V+1$ vector multiplet gauge fields $W^I_\mu$ as
\begin{equation}
	A_{\mu\nu} = \partial_\mu A_\nu - \partial_\nu A_\mu~, \quad F^I_{\mu\nu} = \partial_\mu W^I_\nu - \partial_\nu W^I_\mu~.
\end{equation}
The self-dual and anti-self-dual parts of these field strengths are
\begin{equation}
	A^\pm_{\mu\nu} = \frac{1}{2}\left(A_{\mu\nu} \pm \tilde{A}_{\mu\nu}\right)~,\quad F^{\pm I}_{\mu\nu} = \frac{1}{2}\left(F^I_{\mu\nu} \pm \tilde{F}^I_{\mu\nu}\right)~,
\label{eq:dual}
\end{equation}
where the dual field strengths $\tilde{A}_{\mu\nu}$ and $\tilde{F}^I_{\mu\nu}$ in our conventions are
\begin{equation}
	\tilde{A}_{\mu\nu} = -\frac{i}{2}\varepsilon_{\mu\nu\rho\sigma}A^{\rho\sigma}~,\quad \tilde{F}^I_{\mu\nu} = -\frac{i}{2}\varepsilon_{\mu\nu\rho\sigma}F^{\rho\sigma I}~.
\end{equation}
We denote antisymmetrized and symmetrized indices by
\begin{equation}
	[\mu\nu] = \frac{1}{2}(\mu\nu - \nu\mu)~,\quad (\mu\nu) = \frac{1}{2}(\mu\nu + \nu\mu)~.
\end{equation}

To make it manageable to present equations in the following work we define the composite fields
\begin{equation}\begin{alignedat}{3}
	& \mathcal{F}^{-I}_{\mu\nu} & {} = {} & F^{-I}_{\mu\nu} - \frac{1}{4}\bar{X}^I T^-_{\mu\nu}~, \\
	&\hat{A} & {} = {} & T^-_{\mu\nu}T^{-\mu\nu}~, \\
	&\hat{F}^-_{\mu\nu} & {} = {} & {-16}\left(W_{\mu\nu\rho\sigma}T^{-\rho\sigma} + D T^-_{\mu\nu} + 2i A_{\rho[\mu}T^{-\rho}_{\nu]}\right)~,\\
	& \hat{C} & {} = {} & 32\left( W_{\mu\nu\rho\sigma}W^{\mu\nu\rho\sigma} + i{}^* W_{\mu\nu\rho\sigma}W^{\mu\nu\rho\sigma} + 6 D^2 - 2 A_{\mu\nu}A^{\mu\nu} - 2 A_{\mu\nu}\tilde{A}^{\mu\nu} \right. \\
	&&&\left. - \frac{1}{2}T^{-\mu\nu} D_\mu D^\rho T^+_{\rho\nu} + \frac{1}{4}R^\mu_{~\nu}T^-_{\mu\rho}T^{+\nu\rho} + \frac{1}{256}T^-_{\mu\nu}T^{-\mu\nu}T^+_{\rho\sigma}T^{+\rho\sigma} \right)~,
	\label{eqn:compositedefs}
\end{alignedat}\end{equation}
where the dual to the Weyl tensor is
\begin{equation}
	{}^*W_{\mu\nu\rho\sigma} = \frac{1}{2}\varepsilon_{\mu\nu}^{~~\lambda\tau}W_{\rho\sigma\lambda\tau}~.
\end{equation}
The composite fields have significance in the underlying superconformal multiplet calculus. However, in this paper we take a low-brow attitude where they represent nothing but notation for combinations of fundamental fields, both physical and auxiliary. 

We define the supercovariant derivative $D^\mu$ which acts on a field $\phi$ with chiral weight $c$ by
\begin{equation}
	D^\mu \phi = \left(\nabla^\mu + i c A^\mu\right)\phi~,
\label{eq:superderiv}
\end{equation}
where $\nabla^\mu$ is the ordinary covariant derivative.  The only (non-composite) fields with non-zero chiral weights are the scalars $X^I$ and the anti-self-dual tensor $T^-_{\mu\nu}$.  The fields $X^I$ and $T^-_{\mu\nu}$ have chiral weight $c = -1$, while their Hermitian conjugates $\bar{X}^I$ and $T^+_{\mu\nu}$ have the opposite chiral weight $c = +1$.  The supercovariant derivative acts on these fields via
\begin{equation}
	D^\mu X^I = (\nabla^\mu - i A^\mu)X^I~, \quad D^\mu T^-_{\rho\sigma} = (\nabla^\mu - i A^\mu)T^-_{\rho\sigma}~.
\end{equation}
The scalar operators $D^\mu D_\mu$ and $\nabla^\mu \nabla_\mu$ are both useful. They are distinguished by the notation
\begin{equation}
	\square = D^\mu D_\mu~,\quad \nabla^2 = \nabla^\mu \nabla_\mu~.
\end{equation}

To summarize, we present all of the fields and their corresponding chiral weight $c$ (which determines how the supercovariant derivative (\ref{eq:superderiv}) acts on the field) in table~\ref{table:sugrafields}.  We will need to find the equations of motion for all fundamental fields, both physical and auxiliary, but not the composite fields; those are defined for notational reasons only.
\bgroup
\def\arraystretch{1.5}
\begin{table}[H]
\centering
\begin{tabular}{|c|c|c|c|c|c|c||c|c|c|c|}\cline{2-11}
	\multicolumn{1}{c|}{} & \multicolumn{6}{c||}{Fundamental}&\multicolumn{4}{c|}{Composite} \\ \hline
	 Field & $g_{\mu\nu}$ & $W_\mu^I$ & $X^I$ & $A_\mu$ & $T^-_{\mu\nu}$ & $D$ & $\mathcal{F}^-_{\mu\nu}$ & $\hat{A}$ & $\hat{F}^-_{\mu\nu}$ & $\hat{C}$ \\ \hline
	 Chiral weight & $0$ & $0$ & $-1$ & $0$ & $-1$ & $0$ & $0$ & $-2$ & $-1$ & $0$ \\ \hline
\end{tabular}
\caption{Summary of the fields (and their corresponding chiral weight $c$) in our theory. The conjugate fields have opposite chiral weights.}
\label{table:sugrafields}
\end{table}
\egroup

%%%%%%%%%%%%%%%%%%%%%%%%%%%%%%%%%%%%%%%%%%%%%%%%%%%%%%%%%%%%%%%%%%

\subsection{Prepotential}
\label{subsec:sugra_prepotential}

%%%%%%%%%%%%%%%%%%%%%%%%%%%%%%%%%%%%%%%%%%%%%%%%%%%%%%%%%%%%%%%%%%

The interactions of $\mathcal{N}=2$ supergravity coupled to vector multiplets can be specified succinctly by a prepotential~\cite{deWit:1980lyi,deWit:1983xhu,deWit:1984rvr}. In the two-derivative theory, the prepotential is a meromorphic function of the complex scalars $X^I$.  A large class of higher-derivative corrections can be incorporated by considering generalized prepotentials that are functions of $\hat{A} = T^-_{\mu\nu}T^{-\mu\nu}$ as well.  We will denote the prepotential by
\begin{equation}
	F \equiv F(X^I,\hat{A})~.
\end{equation}

The derivatives of the prepotential are denoted
\begin{equation}
	\frac{\partial F}{\partial X^I} = F_I~,\quad \frac{\partial F}{\partial\hat{A}} = F_A~.
\end{equation}
The prepotential is holomorphic, so 
\begin{equation}
	F_{\bar{I}} = F_{\bar{A}} = 0
\label{eq:prepot_holo}
\end{equation}

The prepotential is homogeneous of degree two under weighted Weyl transformations where the scalar fields $X^I$ and $\hat{A} = T^-_{\mu\nu}T^{-\mu\nu}$ have Weyl weight $w = 1$ and $w = 2$, respectively.  Thus, the prepotential satisfies the homogeneity relation
\begin{equation}
	F_I X^I + 2 F_A \hat{A} = 2 F~.
\label{eq:prepot_homog}
\end{equation}

%%%%%%%%%%%%%%%%%%%%%%%%%%%%%%%%%%%%%%%%%%%%%%%%%%%%%%%%%%%%%%%%%%

\subsection{Action}
\label{subsec:sugra_action}

%%%%%%%%%%%%%%%%%%%%%%%%%%%%%%%%%%%%%%%%%%%%%%%%%%%%%%%%%%%%%%%%%%

We can now present the bosonic part of the $\mathcal{N}=2$ supergravity action as
\begin{equation}
	\mathcal{S} = \int d^4x\,\sqrt{-g}\,\mathcal{L}~,
\end{equation}
with
\begin{equation}\begin{aligned}
	8\pi\mathcal{L} &= -\frac{i}{2} (F_I \bar{X}^I - \bar{F}_I X^I)R + i D^\mu F_I D_\mu \bar{X}^I + \text{h.c.} \\
	&\quad + \bigg{[}\frac{i}{4}F_{IJ}\mathcal{F}^{-I}_{\mu\nu}\mathcal{F}^{-\mu\nu J} - \frac{i}{8} F_I \mathcal{F}^{+I}_{\mu\nu} T^{+\mu\nu} - \frac{i}{32}F T_{\mu\nu}^+ T^{+\mu\nu} \\
	&\quad + \frac{i}{2}F_{AI}\mathcal{F}^{-I}_{\mu\nu}\hat{F}^{-\mu\nu} + \frac{i}{2} F_A \hat{C} + \frac{i}{4}F_{AA} \hat{F}_{\mu\nu}^- \hat{F}^{-\mu\nu} \bigg{]} +\text{h.c.}~,
\label{eq:lsimple}
\end{aligned}\end{equation}
where $\mathcal{F}^{- I}_{\mu\nu}$, $\hat{A}$, $\hat{F}^-_{\mu\nu}$ and $\hat{C}$ are the composite fields defined in
(\ref{eqn:compositedefs}), and $F = F(X^I,\hat{A})$ is the prepotential discussed in section~\ref{subsec:sugra_prepotential}. Any solution to the equations of motion of this action must also be subject to the constraint
\begin{equation}
	D = -\frac{1}{3}R~,
\label{eq:dconstraint}
\end{equation}
which arises from making sure that the auxiliary $D$-field equation of motion is consistent with the other equations of motion.  The details of how we arrived at the Lagrangian (\ref{eq:lsimple}) are given in the appendix.

The coefficient of the Ricci scalar in the action is determined by the  K\"{a}hler potential
\begin{equation}
	e^{-\mathcal{K}} \equiv i\left(F_I \bar{X}^I - \bar{F}_I X^I\right)~.
\end{equation}
At face value this means the metric is in a non-canonical frame since the Ricci scalar normalization depends on the fields $X^I$ and $\hat{A}$.  However, the theory is invariant under a local Weyl symmetry that acts as a gauge symmetry and constrains the scalars $X^I$ such that only $n_V$ of them are independent. In particular, we can gauge-fix our theory and choose one of the scalars such that the K\"{a}hler potential is constant. The low-energy action will then reduce to an Einstein-Hilbert action coupled to matter.

%%%%%%%%%%%%%%%%%%%%%%%%%%%%%%%%%%%%%%%%%%%%%%%%%%%%%%%%%%%%%%%%%%

\section{Minimal Supergravity with (Weyl)$^2$ Corrections}
\label{sec:minsugra}

%%%%%%%%%%%%%%%%%%%%%%%%%%%%%%%%%%%%%%%%%%%%%%%%%%%%%%%%%%%%%%%%%%

In this section we specialize to minimal supergravity, where gravity is coupled to a single vector field, with higher-derivative corrections in the form of a supersymmetrized (Weyl)$^2$ term.  We will present the prepotential and action for the theory and derive the full equations of motion.

%%%%%%%%%%%%%%%%%%%%%%%%%%%%%%%%%%%%%%%%%%%%%%%%%%%%%%%%%%%%%%%%%%

\subsection{Prepotential and Action}

%%%%%%%%%%%%%%%%%%%%%%%%%%%%%%%%%%%%%%%%%%%%%%%%%%%%%%%%%%%%%%%%%%

Following the discussion in section~\ref{subsec:sugra_whatwedo}, the field content for a theory with $n_V=0$ physical $\mathcal{N}=2$ vector multiplets is as follows.

There is a Weyl multiplet containing the metric $g_{\mu\nu}$ and a single vector multiplet containing a physical $U(1)$ gauge field $W_\mu$ and a complex scalar $X$.  The complex scalar field will eventually be gauge-fixed, leaving no physical scalars. The off-shell formalism reviewed in section~\ref{sec:sugra} (and the appendix) further requires that our theory contain the auxiliary $U(1)_R$ vector field $A_\mu$, the auxiliary scalar $D$ and the auxiliary antisymmetric tensor $T^-_{\mu\nu}$.  The Lagrangian will be a function of all these fields.

The prepotential in the minimal theory is a function only of the complex scalar $X$ and the composite field $\hat{A} = T^-_{\mu\nu}T^{-\mu\nu}$.  In this paper we focus on four-derivative corrections to minimal supergravity,  which corresponds to a term in the prepotential that is linear in $\hat{A}$. Higher powers of $\hat{A}$ will give rise to corrections with at least six derivatives.  The homogeneity (\ref{eq:prepot_homog}) and holomorphicity (\ref{eq:prepot_holo}) conditions require the prepotential take the form
\begin{equation}
	F(X,\hat{A}) = -\frac{i}{2}X^2 - c \hat{A}~,\quad c = c_1 + i c_2 \in \mathbb{C}~.
\label{eq:pure_prepotential}
\end{equation}
We can now specialize the full bosonic Lagrangian (\ref{eq:lsimple}) to the minimal supergravity case defined by the prepotential (\ref{eq:pure_prepotential}).  Dropping all total derivative terms, we find
\begin{equation}\begin{aligned}
	8\pi\mathcal{L} &= -|X|^2 R + 2 D^\mu X D_\mu \bar{X} +  \frac{1}{4}F_{\mu\nu}F^{\mu\nu} - \frac{1}{4}F_{\mu\nu}\left(X T^{+\mu\nu} + \bar{X}T^{-\mu\nu}\right) \\
	&\quad+ \frac{1}{32}\left(X^2 T^+_{\mu\nu}T^{+\mu\nu} + \bar{X}^2 T^-_{\mu\nu}T^{-\mu\nu} \right) + 32c_2\bigg{(}W_{\mu\nu\rho\sigma}W^{\mu\nu\rho\sigma} + 6 D^2 \\
	&\quad - 2 A_{\mu\nu}A^{\mu\nu} + \frac{1}{2}(D_\mu T^{-\mu\nu})(D^\rho T^+_{\rho\nu}) + \frac{1}{4}R^\mu_{~\nu}T^-_{\mu\rho}T^{+\nu\rho} \\
	&\quad + \frac{1}{512}T^-_{\mu\nu}T^{-\mu\nu}T^+_{\rho\sigma}T^{+\rho\sigma}  \bigg{)}~.
\label{eq:lpure_action}
\end{aligned}\end{equation}
As we discussed in the general case, any solution is also subject to the constraint equation $D = -\frac{1}{3}R$.

The coefficient of the Ricci scalar is determined by the complex scalar $X$.  As we noted in section~\ref{subsec:sugra_action}, the local Weyl symmetry of the action allows a gauge where $X$ is an arbitrary constant. We will eventually assign it the conventional numerical value but for now we keep $X$ as an independent field.

For $c_2=0$ our minimal $\mathcal{N}=2$ supergravity Lagrangian (\ref{eq:lpure_action}) reduces to the standard two-derivative minimal supergravity, albeit presented in a somewhat unfamiliar form. The new terms are collected in the bracket preceded by the factor $32c_2$. They include first of all an explicit $W_{\mu\nu\rho\sigma}W^{\mu\nu\rho\sigma}$ term, as we wanted, but there are many other terms as well. We interpret the entire expression proportional to $c_2$ as the $\mathcal{N}=2$ supersymmetric completion of $W_{\mu\nu\rho\sigma}W^{\mu\nu\rho\sigma}$.

In the off-shell formalism the auxiliary field $T^-_{\mu\nu}$ is an antisymmetric tensor, a fundamental field. From this point of view the supersymmetric partners of $W_{\mu\nu\rho\sigma}W^{\mu\nu\rho\sigma}$ all contain at most two derivatives. This presents a conceptual advantage because it simplifies the initial value problem. On the other hand, in the context of explicit solutions $T^-_{\mu\nu}$ will coincide with a gauge field strength, with one derivative acting on a gauge field. We will additionally take $D^2=A_{\mu\nu}A^{\mu\nu}=0$ consistently. Therefore the supersymmetric partners of $W_{\mu\nu\rho\sigma}W^{\mu\nu\rho\sigma}$ will all represent four-derivative terms on-shell. 

The coefficient $c_2$ was introduced as the imaginary part of the coupling constant $c=c_1 + i c_2$ in the prepotential (\ref{eq:pure_prepotential}).  All dependence on the real part $c_1$ has dropped out, because $c_1$ couples only to total-derivative terms such as the Chern-Pontryagin terms ${}^*W_{\mu\nu\rho\sigma} W^{\mu\nu\rho\sigma}$ and $A_{\mu\nu}\tilde{A}^{\mu\nu}$. We omitted such terms from the Lagrangian since they do not contribute to the equations of motion.

%%%%%%%%%%%%%%%%%%%%%%%%%%%%%%%%%%%%%%%%%%%%%%%%%%%%%%%%%%%%%%%%%%

\subsection{Equations of Motion}
\label{subsec_eom}

%%%%%%%%%%%%%%%%%%%%%%%%%%%%%%%%%%%%%%%%%%%%%%%%%%%%%%%%%%%%%%%%%%

Many previous studies focused on BPS solutions that preserve the full $\mathcal{N}=2$ supersymmetry, or at least $\frac{1}{2}$-BPS solutions that preserve a residual $\mathcal{N}=1$ supersymmetry.  Such solutions are greatly constrained by relatively simple BPS equations and so it is sufficient to consider a small subset of the equations of motion.  We are interested in solutions that explicitly break supersymmetry, and so we need to derive and solve the full equations of motion for the Lagrangian (\ref{eq:lpure_action}).

The only $D$-dependence in the Lagrangian is the $D^2$ term, and so the $D$-equation of motion forces $D = 0$.  When combined with the constraint equation (\ref{eq:dconstraint}), this forces us to consider solutions with vanishing Ricci scalar
\begin{equation}
	R = 0~.
\label{eq:dconstraint_2}
\end{equation}
We compute the equations of motion for the matter fields $X$, $T^-_{\mu\nu}$, $W_\mu$, and $A_\mu$ to be, respectively,
\begin{equation}\begin{aligned}
	0 &= \square \bar{X} + \frac{1}{2}\bar{X}R + \frac{1}{8}\left(F^+_{\mu\nu} - \frac{1}{4}X T^+_{\mu\nu}\right)T^{+\mu\nu}~,\\
	0 &= \bar{X}\left(F^-_{\mu\nu} - \frac{1}{4}\bar{X}T^-_{\mu\nu}\right) -\frac{c_2}{2}\left(128 D_{[\mu}D^\rho T^+_{\nu]\rho} +  T^-_{\mu\nu}T^+_{\rho\sigma}T^{+\rho\sigma} - 64 R^\rho_{~[\mu}T^+_{\nu]\rho}\right)~, \\
	0 &= D^\mu\left(F^+_{\mu\nu} + F^-_{\mu\nu} - \frac{1}{2}XT^+_{\mu\nu} - \frac{1}{2}\bar{X}T^-_{\mu\nu}\right)~, \\
	0 &= X D^\mu \bar{X} - \bar{X}D^\mu X + 8 c_2\left(T^{-\mu\nu}D^\rho T^+_{\rho\nu} - T^{+\mu\nu}D^\rho T^-_{\rho\nu} - 16i D_\nu A^{\mu\nu} \right)~.
\label{eq:matter_eom}
\end{aligned}\end{equation}
The field strength $F_{\mu\nu}$ must also satisfy the Bianchi identity $D^\mu \tilde{F}_{\mu\nu} = 0$ which we express as
\begin{equation}
	D^\mu\left(F^+_{\mu\nu} - F^-_{\mu\nu}\right) = 0~.
\label{eq:bianchi}
\end{equation}
In order to derive the Einstein equation, we first rewrite the minimal supergravity Lagrangiann (\ref{eq:lpure_action}) as
\begin{equation}
	\mathcal{L} = -\frac{1}{8\pi}|X|^2 R + \mathcal{L}^{(2)} + \mathcal{L}^{(4)}~,
\end{equation}
where $\mathcal{L}^{(2)}$ is the Lagrangian for the two-derivative matter terms
\begin{equation}\begin{aligned}
	\mathcal{L}^{(2)} = \frac{1}{8\pi}\bigg{[}&2 D^\mu X D_\mu \bar{X} + \frac{1}{4}F_{\mu\nu}F^{\mu\nu} - \frac{1}{4}F_{\mu\nu}\left(X T^{+\mu\nu} + \bar{X}T^{-\mu\nu}\right) \\
	&+ \frac{1}{32}\left(X^2 T^+_{\mu\nu}T^{+\mu\nu} + \bar{X}^2 T^-_{\mu\nu}T^{-\mu\nu} \right)\bigg{]}~,
\end{aligned}\end{equation}
while $\mathcal{L}^{(4)}$ contains all of the four-derivative terms present in the supersymmetrized Weyl invariant
\begin{equation}\begin{aligned}
	\mathcal{L}^{(4)} &= \frac{4c_2}{\pi}\left(W_{\mu\nu\rho\sigma}W^{\mu\nu\rho\sigma} + 6 D^2 - 2 A_{\mu\nu}A^{\mu\nu} + \frac{1}{2}(D_\mu T^{-\mu\nu})(D^\rho T^+_{\rho\nu}) \right. \\
	&\quad\quad\quad\quad\quad\left.+ \frac{1}{4}R^\mu_{~\nu}T^-_{\mu\rho}T^{+\nu\rho} + \frac{1}{512}T^-_{\mu\nu}T^{-\mu\nu}T^+_{\rho\sigma}T^{+\rho\sigma}  \right)~.
\end{aligned}\end{equation}
The Einstein equation can now be expressed as
\begin{equation}
	\frac{1}{4\pi}|X|^2\left(R_{\mu\nu} - \frac{1}{2}g_{\mu\nu}R\right) = T_{\mu\nu}^{(2)} + T_{\mu\nu}^{(4)}~,
\label{eq:einstein}
\end{equation}
where $T_{\mu\nu}^{(2)}$ is the energy-momentum tensor for the two-derivative matter
\begin{equation}\begin{aligned}
	T_{\mu\nu}^{(2)} = \frac{2}{\sqrt{-g}}\frac{\delta\left( \sqrt{-g}\mathcal{L}^{(2)}\right)}{\delta g^{\mu\nu}} = \frac{1}{4\pi}\bigg{[}&2(D_\mu X)(D_\nu \bar{X}) - g_{\mu\nu}(D_\rho X)(D^\rho \bar{X}) \\
	&+ F^+_{\mu\rho}F^{-\rho}_\nu - \frac{1}{4}\left(X F^-_{\mu\rho}T^{+\rho}_\nu + \bar{X}F^+_{\mu\rho}T^{-\rho}_\nu\right)\bigg{]}~,
	\label{eq:twoderem}
\end{aligned}\end{equation}
while $T^{(4)}_{\mu\nu}$ is the energy-momentum tensor for the four-derivative parts of the action
\begin{equation}\begin{aligned}
 T^{(4)}_{\mu\nu} &= \frac{2}{\sqrt{-g}}\frac{\delta\left( \sqrt{-g}\mathcal{L}^{(4)}\right)}{\delta g^{\mu\nu}} \\
 &= \frac{8 c_2}{\pi} \left(4 R_{\mu\rho}R_\nu^{~\rho} - g_{\mu\nu}R_{\rho\sigma}R^{\rho\sigma} - \frac{4}{3}R_{\mu\nu}R+\frac{1}{3}g_{\mu\nu}R^2 - 2 \square R_{\mu\nu}\right. \\
	&\quad + 4 D^\rho D_\mu R_{\nu\rho} + \frac{1}{3}g_{\mu\nu}\square R - \frac{4}{3}D_\mu D_\nu R -4 A_{\mu\rho}A_\nu^{~\rho} + g_{\mu\nu}A_{\rho\sigma}A^{\rho\sigma} \\
	&\quad - \frac{1}{4} g_{\mu\nu}(D^\rho T^-_{\rho\tau})(D_\sigma T^{+\sigma\tau}) + \frac{1}{2}(D_\mu T^-_{\nu\rho})(D_\sigma T^{+\sigma\rho}) \\
	&\quad + \frac{1}{2}(D_\mu T^+_{\nu\rho})(D_\sigma T^{-\sigma\rho}) + \frac{1}{2}(D^\rho T^-_{\rho\mu})(D^\sigma T^+_{\sigma\nu}) \\
	&\quad+\frac{1}{1024}g_{\mu\nu}T^-_{\rho\sigma}T^{-\rho\sigma}T^+_{\tau\lambda}T^{+\tau\lambda} - \frac{1}{8}g_{\mu\nu}R_{\rho\sigma}T^-_{\rho\tau}T^{+\tau}_\sigma + \frac{1}{2} R_{\mu\rho}T^-_{\nu\sigma}T^{+\rho\sigma} \\
	&\quad+ \frac{1}{4}R^{\rho\sigma}T^-_{\mu\rho}T^+_{\nu\sigma} + \frac{1}{4}D_\rho D_\mu(T^-_{\nu\sigma}T^{+\rho\sigma}) - \frac{1}{8}\square(T^-_{\mu\rho}T^{+\nu\rho}) \\
	&\quad\left. - \frac{1}{8}g_{\mu\nu} D_\rho D_\sigma(T^{-\rho\tau}T^{+\sigma}_{~~~\tau})\right)~.
	\label{eq:fourderem}
\end{aligned}\end{equation}
In summary, we have shown that any solution to our minimal supergravity theory must satisfy the matter field equations of motion (\ref{eq:matter_eom}), the Bianchi identity (\ref{eq:bianchi}), the Einstein equation (\ref{eq:einstein}), and must have a geometry with vanishing Ricci scalar $R = 0$.

%%%%%%%%%%%%%%%%%%%%%%%%%%%%%%%%%%%%%%%%%%%%%%%%%%%%%%%%%%%%%%%%%%

\section{Non-Supersymmetric Solutions}
\label{sec:em}

%%%%%%%%%%%%%%%%%%%%%%%%%%%%%%%%%%%%%%%%%%%%%%%%%%%%%%%%%%%%%%%%%%

In this section we embed arbitrary solutions to Einstein-Maxwell theory into the minimal $\mathcal{N} = 2$ supergravity theory (with a supersymmetrized (Weyl)$^2$ correction) presented in section~\ref{sec:minsugra}. The matter fields of the higher-derivative gravity are specified in terms of the matter in the Einstein-Maxwell theory. The geometry that supports the Einstein-Maxwell solution is unchanged when considered as solution to higher-derivative gravity. 

%%%%%%%%%%%%%%%%%%%%%%%%%%%%%%%%%%%%%%%%%%%%%%%%%%%%%%%%%%%%%%%%%%

\subsection{Einstein-Maxwell}
\label{subsec:em_einmax}

%%%%%%%%%%%%%%%%%%%%%%%%%%%%%%%%%%%%%%%%%%%%%%%%%%%%%%%%%%%%%%%%%%

The starting point is the standard Einstein-Maxwell theory
\begin{equation}
	\mathcal{L}_\text{EM} =  - \frac{1}{2\kappa^2}\left(\mathbf{R} + \frac{1}{4}\mathbf{F}_{\mu\nu}\mathbf{F}^{\mu\nu}\right)~,
\end{equation}
where $\kappa^2 = 8\pi G_N$.  We are using boldfaced symbols 
$\mathbf{g}_{\mu\nu}$, $\mathbf{R}$ , and $\mathbf{F}_{\mu\nu}$ for the metric, Ricci scalar, and electromagnetic field strength in Einstein-Maxwell theory in order to avoid any confusion with related quantities in the higher-derivative supergravity Lagrangian (\ref{eq:lpure_action}).

Any solution to Einstein-Maxwell theory satisfies the Maxwell equations and the Bianchi identities, which we package together as the Maxwell-Bianchi equations
\begin{equation}
	\nabla^\mu \mathbf{F}^\pm_{\mu\nu} = 0~,
\label{eq:maxbianchi}
\end{equation}
where the self-dual and anti-self-dual parts of the field strength are defined using the conventions in section~\ref{subsec:sugra_definitions}. The geometry and the matter fields are related by the Einstein equation
\begin{equation}
	\mathbf{R}_{\mu\nu} - \frac{1}{2}\mathbf{g}_{\mu\nu}\mathbf{R} = -\mathbf{F}^-_{\mu\rho}\mathbf{F}^{+\rho}_\nu~.
\label{eq:em_ein}
\end{equation}
We are particularly interested in Kerr-Newman black hole solutions but our embedding will apply to 
\emph{any} solution of Einstein-Maxwell theory.

%%%%%%%%%%%%%%%%%%%%%%%%%%%%%%%%%%%%%%%%%%%%%%%%%%%%%%%%%%%%%%%%%%

\subsection{Embedding}
\label{subsec:em_embed}

%%%%%%%%%%%%%%%%%%%%%%%%%%%%%%%%%%%%%%%%%%%%%%%%%%%%%%%%%%%%%%%%%%

Starting from a solution to Einstein-Maxwell theory we specify the matter fields in the higher-derivative theory as 
\begin{equation}\begin{aligned}
	X = \frac{\sqrt{4\pi}}{\kappa}~, \quad A_\mu = 0~, \quad T^\pm_{\mu\nu} = 4\mathbf{F}^\pm_{\mu\nu}~, \quad F^{\pm}_{\mu\nu} = \frac{1}{4}X T^\pm_{\mu\nu} = X \mathbf{F}^\pm_{\mu\nu}~.
\label{eq:em_embed}
\end{aligned}\end{equation}
As mentioned previously, the geometry is unchanged. 

The numerical value of $X$ is such that the Ricci scalar term in the Lagrangian (\ref{eq:lpure_action}) is normalized correctly 
\begin{equation}
	\mathcal{L} = -\frac{1}{2\kappa^2}R + \ldots~.
\end{equation}
By choosing $A_\mu = 0$, the supercovariant derivative operator $D^\mu$ reduces to the ordinary covariant derivative operator $\nabla^\mu$. 

It is rather straightforward to show that all the matter field equations of motion (\ref{eq:matter_eom}) are satisfied by the matter (\ref{eq:em_embed}). Since $\mathbf{F}^\pm_{\mu\nu}$ is divergence-free by the Maxwell-Bianchi equations (\ref{eq:maxbianchi}), $T^\pm_{\mu\nu}$ must be divergence-free as well
\begin{equation}
	D^\mu T^\pm_{\mu\nu} = 0~.
\label{eq:t_divfree}
\end{equation}
Since $X$ is constant and $A_{\mu\nu}=0$ the final equation in (\ref{eq:matter_eom}) follows. We also have $D^\mu F^\pm_{\mu\nu} =0$ (since $X$ is constant) and so the gauge field equations in the third line of (\ref{eq:matter_eom}) are satisfied. The scalar equation of motion is satisfied because $X$ is constant, the geometry has $R=0$, and the matter satisfies 
\begin{equation}
{\cal F}^\pm_{\mu\nu} = F^\pm_{\mu\nu} - {1\over 4} X T^\pm_{\mu\nu}= 0 ~.
\label{eq:FcalZero}
\end{equation}
The equation of motion for the antisymmetric tensor $T^-_{\mu\nu}$ is slightly less obvious. It is satisfied due to the following identities for (anti-)self-dual tensors in 4D:
\begin{equation}
	T^+_{\mu\nu}T^{-\rho\sigma} + T^{+\rho\sigma}T^-_{\mu\nu} = 4 \delta^{[\rho}_{[\mu}T^+_{\nu]\tau}T^{-\sigma]\tau}~,\quad T^+_{\mu\nu}T^{-\mu\nu} = 0~.
\label{eq:tprops}
\end{equation}

At this point we still need to verify the Einstein equation (\ref{eq:einstein}).  It is important to note that the only dependence on $c_2$ is in the four-derivative energy-momentum tensor $T^{(4)}_{\mu\nu}$ and not in any of the two-derivative terms.  Since we claim the embedding works for any value of the constant $c_2$, the two-derivative and four-derivative terms must cancel independently.  The original Einstein equation (\ref{eq:einstein}) therefore becomes two separate equations
\begin{equation}
	\frac{1}{4\pi}|X|^2\left(R_{\mu\nu} - \frac{1}{2} g_{\mu\nu}R\right) = T^{(2)}_{\mu\nu} \quad\text{and}\quad T^{(4)}_{\mu\nu} = 0~.
\label{eq:einstein_2}
\end{equation}
The energy-momentum tensor $T^{(2)}_{\mu\nu}$, given in (\ref{eq:twoderem}), simplifies greatly due to the embedding (\ref{eq:em_embed}). The two-derivative part of the Einstein equations (\ref{eq:einstein_2}) becomes
\begin{equation}
	R_{\mu\nu} - \frac{1}{2}g_{\mu\nu}R = -\mathbf{F}^+_{\mu\rho}\mathbf{F}^{-\rho}_\nu~.
\label{eq:ein_2d}
\end{equation}
We recognize this equation as the original condition on the Einstein-Maxwell geometry (\ref{eq:em_ein}).  Taking the trace of this expression yields
\begin{equation}
	R = 0~,
\end{equation}
as required by the constraint equation (\ref{eq:dconstraint_2}) coming from the auxiliary $D$-field.  

The four-derivative part of the Einstein equations (\ref{eq:einstein_2}), with $T^{(4)}_{\mu\nu}$ given in (\ref{eq:fourderem}), becomes 
\begin{equation}\begin{aligned}
	0 &= 4 R_{\mu\rho}R_\nu^{~\rho} - g_{\mu\nu}R_{\rho\sigma}R^{\rho\sigma} - \frac{4}{3}R_{\mu\nu}R+\frac{1}{3}g_{\mu\nu}R^2 - 2 \nabla^2 R_{\mu\nu} \\
	&\quad + 4 \nabla^\rho \nabla_\mu R_{\nu\rho} + \frac{1}{3}g_{\mu\nu}\nabla^2 R - \frac{4}{3}\nabla_\mu \nabla_\nu R \\
	&\quad+\frac{1}{4}g_{\mu\nu}\mathbf{F}^-_{\rho\sigma}\mathbf{F}^{-\rho\sigma}\mathbf{F}^+_{\tau\lambda}\mathbf{F}^{+\tau\lambda} - 2g_{\mu\nu}R_{\rho\sigma}\mathbf{F}^-_{\rho\tau}\mathbf{F}^{+\tau}_\sigma + 8 R_{\mu\rho}\mathbf{F}^-_{\nu\sigma}\mathbf{F}^{+\rho\sigma} \\
	&\quad+ 4R^{\rho\sigma}\mathbf{F}^-_{\mu\rho}\mathbf{F}^+_{\nu\sigma} + 4\nabla_\rho \nabla_\mu(\mathbf{F}^-_{\nu\sigma}\mathbf{F}^{+\rho\sigma}) - 2\nabla^2(\mathbf{F}^-_{\mu\rho}\mathbf{F}^{+\nu\rho}) \\
	&\quad - 2g_{\mu\nu} \nabla_\rho \nabla_\sigma(\mathbf{F}^{-\rho\tau}\mathbf{F}^{+\sigma}_{~~~\tau})~,
\label{eq:ein_4d}
\end{aligned}\end{equation}
upon insertion of the embedding (\ref{eq:em_embed}). It is not immediately obvious that it is realistic to solve this equation. However,
repeated use of $R_{\mu\nu} = -\mathbf{F}^+_{\mu\rho}\mathbf{F}^{-\rho}_\nu$ in (\ref{eq:ein_4d}) and careful simplification shows that it is in fact satisfied identically.  

In summary, we have verified that our embedding (\ref{eq:em_embed}) generates a solution to the higher-derivative theory for each solution to the original Einstein-Maxwell theory. This result relies on supersymmetry of the theory, as the action we consider is far from arbitrary. However, the solutions do not generally preserve any supersymmetry. 

As a check on these results, we consider the special case of extremal Reissner-Nordstr{\"o}m black holes.  We have verified that the BPS equations derived in~\cite{LopesCardoso:2000fp,LopesCardoso:2000qm} are satisfied by our embedding (\ref{eq:em_embed}) for extremal Reissner-Nordstr{\"o}m geometries.  This is expected, since these geometries are known to be $\frac{1}{2}$-BPS domain walls that interpolate between the $\mathcal{N}=2$ supersymmetric AdS$_2\times S^2$ geometry at the horizon and the $\mathcal{N}=2$ supersymmetric Minkowski spacetime at infinity.

%%%%%%%%%%%%%%%%%%%%%%%%%%%%%%%%%%%%%%%%%%%%%%%%%%%%%%%%%%%%%%%%%%

\subsection{Simplified Lagrangian}

%%%%%%%%%%%%%%%%%%%%%%%%%%%%%%%%%%%%%%%%%%%%%%%%%%%%%%%%%%%%%%%%%%

Having showed that the embedding (\ref{eq:em_embed}) satisfies the fairly complicated equations of motion for minimal supergravity with higher-derivative corrections, it is worth understanding why this is the case. We do so by introducing a simplified effective Lagrangian that captures the same dynamics as the original Lagrangian (\ref{eq:lpure_action}) within the context of our embedding.  

As a first step we can consistently eliminate the auxiliary fields $D$ and $A_\mu$ by setting both to zero at the level of the action.  We then use properties of (anti-)self-dual tensors in 4D (\ref{eq:tprops}) to express the simplified Lagrangian as
\begin{equation}\begin{aligned}
	8\pi\mathcal{L}_\text{trunc} &= - |X|^2 R - \frac{1}{4}F_{\mu\nu}F^{\mu\nu}  +  2\nabla^\mu X \nabla_\mu \bar{X}  + \frac{1}{2}\left(F_{\mu\nu}^+ - \frac{1}{4}X T^+_{\mu\nu}\right)^2 + \text{h.c.} \\
	&\quad + 32c_2\bigg{(}W_{\mu\nu\rho\sigma}W^{\mu\nu\rho\sigma} + \frac{1}{4}R^\mu_{~\nu}T^-_{\mu\rho}T^{+\nu\rho} + \frac{1}{512}T^-_{\mu\nu}T^{-\mu\nu}T^+_{\rho\sigma}T^{+\rho\sigma} \\
	&\quad+ \frac{1}{2}(\nabla_\mu T^{-\mu\nu})(\nabla^\rho T^+_{\rho\nu}) \bigg{)}~.
\label{eq:ltrunc_0}
\end{aligned}\end{equation}
We now want to eliminate the auxiliary fields $X$ and $T^-_{\mu\nu}$ from the action by replacing them with their ansatz in the embedding (\ref{eq:em_embed}):
\begin{equation}
	X = \frac{\sqrt{4\pi}}{\kappa}~,\quad T^-_{\mu\nu} = \frac{4}{X}F^-_{\mu\nu}~,
\label{eq:xtansatz}
\end{equation}
at the level of the action.  We can see from (\ref{eq:ltrunc_0}) that $X$ is sourced by the Ricci scalar, which vanishes for Einstein-Maxwell backgrounds, and $F^+_{\mu\nu} - \frac{1}{4}XT^+_{\mu\nu}$, which vanishes in (\ref{eq:xtansatz}).  Similarly, $T^-_{\mu\nu}$ is sourced by $F^-_{\mu\nu} - \frac{1}{4}\bar{X}T^-_{\mu\nu}$ and various other higher-derivative terms that are independent of $F_{\mu\nu}$ and vanish for Einstein-Maxwell backgrounds.  {The elimination (\ref{eq:xtansatz}) is therefore consistent with the $X$ and $T^-_{\mu\nu}$ equations of motion and can be implemented at the level of the action.

To make the normalization simpler we also rescale the vector multiplet field strength by
\begin{equation}
	F_{\mu\nu} \to \frac{\sqrt{4\pi}}{\kappa}F_{\mu\nu}~. 
\end{equation}
After these simplifications we find
\begin{equation}\begin{aligned}
	\mathcal{L}_\text{trunc} &= -\frac{1}{2\kappa^2}\left(R + \frac{1}{4}F_{\mu\nu}F^{\mu\nu}\right) + \frac{4 c_2}{\pi}\bigg{(} W_{\mu\nu\rho\sigma}W^{\mu\nu\rho\sigma} + 4 R^\mu_{~\nu}F^-_{\mu\rho}F^{+\nu\rho} \\
	&\quad + \frac{1}{2}F^-_{\mu\nu}F^{-\mu\nu}F^+_{\rho\sigma}F^{+\rho\sigma} + 8(\nabla_\mu F^{-\mu\nu})(\nabla^\rho F^+_{\rho\nu})\bigg{)}~.
\label{eq:ltrunc_1}
\end{aligned}\end{equation}
This form of the Lagrangian expresses the intuitive notion that our theory is ordinary Einstein-Maxwell theory with addition of a supersymmetrized Weyl invariant that includes mixings between the electromagnetic field strength and the Riemann tensor.  Any solution to the truncated theory (\ref{eq:ltrunc_1}) will automatically be a solution to the minimal supergravity theory (\ref{eq:lpure_action}).  

Our black hole solutions imply that the supersymmetrized Weyl invariant
\begin{equation}\begin{aligned}
	\mathcal{L}_{\rm {\cal N}=2~Weyl} &= W_{\mu\nu\rho\sigma}W^{\mu\nu\rho\sigma} + 4 R^\mu_{~\nu}F^-_{\mu\rho}F^{+\nu\rho} + \frac{1}{2}F^-_{\mu\nu}F^{-\mu\nu}F^+_{\rho\sigma}F^{+\rho\sigma} \\
	&\quad + 8(\nabla_\mu F^{-\mu\nu})(\nabla^\rho F^+_{\rho\nu})
\label{eq:susyweyl_precise}
\end{aligned}\end{equation}
can be included into the Einstein-Maxwell action without consequence to the geometry or the field strength. To understand this claim we rewrite $W_{\mu\nu\rho\sigma}W^{\mu\nu\rho\sigma}$ in terms of the Gauss-Bonnet density $E_4$ as
\begin{equation}
	W_{\mu\nu\rho\sigma}W^{\mu\nu\rho\sigma} = E_4 + 2 R_{\mu\nu}R^{\mu\nu} - \frac{2}{3}R^2~,
\end{equation}
and find
\begin{equation}
	\mathcal{L}_{\rm {\cal N}=2~Weyl} = E_4 + 2\left(R_{\mu\nu}+F^-_{\mu\rho}F_\nu^{+\rho} \right)^2 - \frac{2}{3}R^2 + 8(\nabla_\mu F^{-\mu\nu})(\nabla^\rho F^+_{\rho\nu})~.
\label{eq:ltrunc_2}
\end{equation}
The Gauss-Bonnet density $E_4$ does not contribute to the equations of motion because it is topological.  The remaining terms $(R_{\mu\nu}+F^-_{\mu\rho}F_\nu^{+\rho})^2$, $R^2$, and $(\nabla_\mu F^{-\mu\nu})(\nabla^\rho F^+_{\rho\nu})$ are all quadratic in expressions that vanish for Einstein-Maxwell backgrounds. That explains why these terms can be introduced in the Einstein-Maxwell action without changing the original solutions.  

The simplifications we find are predicated on the precise combination of four-derivative terms appearing in (\ref{eq:susyweyl_precise}); any others would lead to complicated corrections of the solutions (see {\it e.g.}~\cite{Kats:2006xp,Lu:2015cqa}).
In our context those coefficients were dictated by the $\mathcal{N} = 2$ supersymmetry of the theory. Thus supersymmetry is responsible for substantial simplifications even for solutions that do not preserve any supersymmetry. 

It was previously noticed in~\cite{Sen:2005iz} that the entropy of supersymmetric black holes in heterotic string theory is the same whether one introduces higher-derivative corrections in the form of a supersymmetrized Weyl invariant or an ordinary Gauss-Bonnet term.  This led to arguments (see \emph{e.g.}~\cite{Sen:2011ba}) that the supersymmetrized Weyl invariant should coincide with the Gauss-Bonnet density on-shell.  Our supersymmetrized Weyl invariant (\ref{eq:ltrunc_2}) makes this argument concrete.  This is particularly surprising in the near-horizon region of BPS black holes: the AdS$_2\times S^2$ geometry has vanishing Weyl tensor, yet the supersymmetrized Weyl invariant is non-zero and matches the Gauss-Bonnet density exactly.

%%%%%%%%%%%%%%%%%%%%%%%%%%%%%%%%%%%%%%%%%%%%%%%%%%%%%%%%%%%%%%%%%%

\section{Properties of Black Holes in Higher-Derivative Gravity}
\label{sec:properties}

%%%%%%%%%%%%%%%%%%%%%%%%%%%%%%%%%%%%%%%%%%%%%%%%%%%%%%%%%%%%%%%%%%

In this section we analyze properties of Kerr-Newman black holes considered as solutions to minimal supergravity with higher-derivative corrections. We show that the black hole entropy simplifies when the theory has $\mathcal{N}=2$ supersymmetry.

\subsection{Black Hole Entropy}
The black hole entropy in the higher-derivative theory is given by the Wald entropy formula~\cite{Wald:1993nt,Iyer:1994ys,Jacobson:1993vj}.  The entropy is
\begin{equation}
	S_\text{Wald} = 2\pi \int_H \frac{\delta \mathcal{L}}{\delta R_{\mu\nu\rho\sigma}} \epsilon_{\mu\nu} \epsilon_{\rho\sigma} \sqrt{h}\, d^2x~,
\label{eq:wald}
\end{equation}
where $h_{ij}$ is the induced metric on the black hole horizon $H$ and $\epsilon_{\mu\nu}$ is the (antisymmetric) unit binormal to the horizon, normalized such that $\epsilon_{\mu\nu}\epsilon^{\mu\nu} = -2$.  
Four-derivative terms in the action give rise to an integrand that includes terms linear in the curvature and terms with two derivatives acting on the matter fields. Each of these terms in the integrand is somewhat intricate and upon integration they will generally give complicated contributions to the entropy. 

However, $\mathcal{N}=2$ supersymmetry dictates relations between the coefficients of these contributions such that the four-derivative terms combine into the expression (\ref{eq:ltrunc_2}).  Any part of the action that is quadratic in terms that vanish on-shell cannot contribute to the Wald entropy (\ref{eq:wald}), since the entropy is determined by a linear variation. For the purposes of computing the Wald entropy it is therefore sufficient to add the Gauss-Bonnet term 
\begin{equation}
	\mathcal{L}_\text{GB} = \frac{4c_2}{\pi}E_4
\end{equation}
to the standard Einstein-Hilbert Lagrangian. This is a considerable simplification. 

The contribution to the Wald entropy from a Gauss-Bonnet term in the action has been studied in detail~\cite{Jacobson:1993xs}. It is known to be purely topological, depending only on the Euler characteristic of the horizon. The total Wald entropy, including the area law due to the Einstein-Hilbert action, becomes
\begin{equation}
	S_\text{Wald} = \frac{A_H}{4 G_N} + 128\pi\chi_{(2)}c_2~,
\label{eq:entropy}
\end{equation}
where $\chi_{(2)}$ is the Euler characteristic of the black hole horizon\footnote{Our curvature conventions are set by the sign on the Ricci scalar in the Einstein-Hilbert action. The curvature of a sphere is negative and the Euler character (\ref{eq:horizon_euler}) has an unusual overall minus sign.} 
\begin{equation}
	\chi_{(2)} = -\frac{1}{4\pi}\int_H dA\,R_{(2)} ~.
\label{eq:horizon_euler}
\end{equation}
For general Kerr-Newman black holes, the Euler characteristic of the horizon is $\chi_{(2)} = 2$, and so the Wald entropy (\ref{eq:entropy}) becomes
\begin{equation}
	S_\text{Wald} = \frac{A_H}{4 G_N} + 256\pi c_2~.
\label{eq:entropy_kn}
\end{equation}
This is the entropy of a Kerr-Newman black hole, including the higher-derivative correction in the form of a supersymmetrized Weyl invariant.  

In the special case of vanishing charge, the black hole geometry is Ricci flat $R_{\mu\nu} = 0$ and so it is obvious that the Weyl invariant coincides with the Gauss-Bonnet term on-shell. We find that this well-known statement generalizes to Kerr-Newman black holes. That is interesting because this family includes a BPS limit, where the black hole preserves the supersymmetry of the theory and the microscopic description is under control. Previous studies~\cite{Castro:2008ne,Castro:2007hc,Castro:2007ci,Mohaupt:2000mj,LopesCardoso:2000qm,LopesCardoso:1998tkj,Sen:2005iz,LopesCardoso:1999cv,LopesCardoso:1999fsj,LopesCardoso:1999xn,Sahoo:2006rp,Alishahiha:2006jd,Kraus:2005vz} have found that higher-derivative corrections in string theory gives rise to a correction of the form (\ref{eq:entropy_kn}) with a numerical coefficient that can be matched with microscopic considerations. 

Our result for the correction to the black hole entropy (\ref{eq:entropy_kn}) has no dependence whatsoever on the parameters of the black hole. The deformation away from the BPS limit by adding mass and introducing angular momentum does not change the correction due to higher-order derivatives. This is reminiscent of our previous result~\cite{Charles:2015eha} that quantum corrections to Kerr-Newman black holes are universal and similarly insensitive to deformations off extremality. For both classes of corrections it is significant that the theory preserves $\mathcal{N}=2$ supersymmetry but it is unimportant whether the black holes preserve the supersymmetry of the theory.

%%%%%%%%%%%%%%%%%%%%%%%%%%%%%%%%%%%%%%%%%%%%%%%%%%%%%%%%%%%%%%%%%%

\subsection{OSV Conjecture}

%%%%%%%%%%%%%%%%%%%%%%%%%%%%%%%%%%%%%%%%%%%%%%%%%%%%%%%%%%%%%%%%%%

The correction to the entropy due to the higher-derivative terms is just a constant, independent of the black hole parameters. The value of the constant is therefore captured by the BPS limit and so it can be interpreted in string theory, {\it e.g.} following the OSV conjecture~\cite{Ooguri:2004zv}. 

For extremal BPS black holes, the attractor mechanism~\cite{Ferrara:1995ih,Ferrara:1996dd,Strominger:1996kf,Sen:2005wa} specifies scalars in the horizon AdS$_2\times S^2$ geometry in terms of the charges $(p^I,q_I)$ by the attractor equations
\begin{equation}
	p^I = \text{Re}[CX^I]~,
\label{eq:attract_p}
\end{equation}
\begin{equation}
	q_I = \text{Re}[CF_I]~,
\label{eq:attract_q}
\end{equation}
where $C$ is an arbitrary scaling parameter chosen as
\begin{equation}
	C^2 \hat{A} = 256~,
\label{eq:cdef}
\end{equation}
with $\hat{A}$ evaluated at the horizon. Expressing the real and imaginary parts of the scalars as
\begin{equation}
	C X^I = p^I + \frac{i}{\pi}\phi^I~,
\end{equation}
the black hole potential is
\begin{equation}
\label{eq:bh_potential}
	\mathcal{F}(\phi^I,p^I) = -\pi\text{ Im}[C^2 F(X^I,\hat{A})]~,
\end{equation}
in a mixed ensemble defined as a microcanonical ensemble of magnetic charges $p^I$ and a canonical ensemble of electric charges $q_I$ with chemical potentials $\phi^I$. The black hole entropy, including higher-derivative terms, is then given by the Legendre transform
\begin{equation}
	S_\text{BH}(q_I,p^I) = \left(1 - \phi^I \frac{\partial}{\partial \phi^I}\right)\mathcal{F}(\phi^I,p^I)~,
\end{equation}
where the electric potentials $\phi^I$ have been eliminated in favor of the electric charges $q_I$ through the attractor equation (\ref{eq:attract_q}).

In the case of our minimal prepotential (\ref{eq:pure_prepotential}) the attractor equations are
\begin{equation}
	p = \text{Re}[C X]~,\quad q = \text{Im}[C X] = \frac{1}{\pi}\phi~,
\end{equation}
and the black hole potential (\ref{eq:bh_potential}) becomes
\begin{equation}
\mathcal{F}(\phi,p) = \frac{\pi}{2}p^2 - \frac{1}{2\pi}\phi^2 + 256\pi c_2~.
\end{equation}
The Legendre transform of this potential gives the black hole entropy 
\begin{equation}
	S_\text{BH} =   \frac{\pi}{2}(q^2 + p^2) + 256\pi c_2~.
\end{equation}
The first term agrees with the classical area law for an extremal Reissner-Nordstr\"{o}m black hole with dyonic $U(1)$ charge, and the correction agrees with our result (\ref{eq:entropy_kn}) computed using the Wald entropy formalism.  

The OSV conjecture~\cite{Ooguri:2004zv} makes connection with microscopic considerations through the relation
\begin{equation}
	Z_\text{BH} = |Z_\text{top}|^2~,
\end{equation}
where $Z_\text{BH}$ is the supersymmetric partition function
\begin{equation}
	Z_\text{BH}(\phi,p) = \text{exp}\big{[}\mathcal{F}(\phi,p)\big{]}
\end{equation}
of a four-dimensional BPS black hole in the mixed ensemble. The partition function of the topological string is similarly
\begin{equation}
Z_\text{\rm top}(\phi,p) = \text{exp}\big{[}\mathcal{F}_{\rm top}(\phi,p)\big{]}~, 
\end{equation}
with 
\begin{equation}
\label{eqn:topstr}
{\cal F}_{\rm top} (\lambda,X)= \sum_{g=0} \lambda_{\rm top}^{2g-2} F_{{\rm top},g}(X)~,
\end{equation}
a perturbative expansion in the coupling constant $\lambda_{\rm top} = {4\pi i \over p + i q}$. The correction we consider is charge-independent, corresponding to the torus partition function with genus $g=1$. 

The OSV conjecture and its possible extensions have been subject to much study and debate,  including~\cite{Dabholkar:2004dq,Dabholkar:2005by,Shih:2005he,Kraus:2005vz,Denef:2007vg}. Since the minimal model we consider has $n_V=0$ moduli it corresponds to a somewhat singular limit, the case of a rigid Calabi-Yau (in the language of the A-model). It would be interesting to study this special case further.

%%%%%%%%%%%%%%%%%%%%%%%%%%%%%%%%%%%%%%%%%%%%%%%%%%%%%%%%%%%%%%%%%%

\section{Discussion}
\label{sec:discussion}

%%%%%%%%%%%%%%%%%%%%%%%%%%%%%%%%%%%%%%%%%%%%%%%%%%%%%%%%%%%%%%%%%%
The motivation for studying Kerr-Newman black holes in string theory is the hope that a precision understanding can be achieved in this setting. We are still far from that goal but we can make some observations in the spirit of phenomenology.

The classical black hole entropy of Kerr-Newman black holes computed from the outer and inner horizons is
\begin{equation}
S_\pm  = 2\pi  \left( (M^2 - {1\over 2}Q^2) \pm \sqrt{ M^2(M^2 - Q^2)  - J^2} \right)~.
\end{equation}
An appealing (but speculative) interpretation of these formulae identifes the combinations
\begin{equation}
	S_{R} = {1\over 2} (S_+ + S_-)~, \quad S_{L} = {1\over 2} (S_+ - S_-)~,
\label{eq:srl}
\end{equation}
with the entropy of factorized right- and left-moving excitations of an underlying CFT with $(0,4)$ supersymmetry~\cite{Larsen:1997ge,Cvetic:1997uw,Cvetic:2009jn}. This theory would be a generalization of the MSW CFT describing the BPS and near-BPS limits~\cite{Maldacena:1997de}. The assignment of supersymmetry is such that the dependence on the angular momentum quantum number can be entirely accounted for by an $SU(2)_R$ current, arbitrarily far from extremality. This is analogous to the standard BMPV model of rotating BPS black holes in five dimensions~\cite{Breckenridge:1996is,Breckenridge:1996sn}. 

The correction to the black hole entropy due to higher-derivative terms (\ref{eq:entropy_kn}) is not just independent of black hole parameters; it is the same when computed at the outer and the inner horizons~\cite{Castro:2013pqa}. Therefore, the prescription (\ref{eq:srl}) with higher-derivative corrections included identifies the corrections as pertaining to the ``Right'' sector, with no corrections in the ``Left'' sector. 

The ``Left'' sector contains the novel excitations, the ones that BPS conditions force into their ground state. These are also the ones that carry the angular momentum of the black hole so the BPS limit is incompatible with rotation. The independence of corrections on black hole parameters suggest that this sector receives no string corrections in the leading approximation. At the level of a phenomenological model this is not unreasonable since, after all, the ``Left'' sector is subject to ${\mathcal N}=4$ supersymmetry, albeit spontaneously broken by the state. 

It would clearly be interesting to develop such a model in more detail.

%%%%%%%%%%%%%%%%%%%%%%%%%%%%%%%%%%%%%%%%%%%%%%%%%%%%%%%%%%%%%%

\acknowledgments

We thank Pedro Lisb{\~a}o and Daniel R. Mayerson for useful discussions and encouragement.  This work was supported in part by the U.S. Department of Energy under grant DE-FG02-95ER40899.

%%%%%%%%%%%%%%%%%%%%%%%%%%%%%%%%%%%%%%%%%%%%%%%%%%%%%%%%%%%%%%

\appendix

%%%%%%%%%%%%%%%%%%%%%%%%%%%%%%%%%%%%%%%%%%%%%%%%%%%%%%%%%%%%%%%%%%

\section{Off-Shell $\mathcal{N}=2$ Supergravity}

%%%%%%%%%%%%%%%%%%%%%%%%%%%%%%%%%%%%%%%%%%%%%%%%%%%%%%%%%%%%%%%%%%

In this section, we summarize the construction of $\mathcal{N}=2$ supergravity in 4D following the off-shell formalism studied in~\cite{LopesCardoso:2000qm, Mohaupt:2000mj,deWit:1980lyi,deWit:1983xhu,deWit:1984rvr}.  We review the bosonic field content and discuss actions that realize the full $\mathcal{N}=2$ supersymmetry with higher-derivative corrections present.  These steps justify the Lagrangian (\ref{eq:lsimple}) that we use to study non-supersymmetric solutions of higher-derivative supergravity.
 
\subsection{$\mathcal{N}=2$ Supergravity Multiplets}

The first step in constructing off-shell $\mathcal{N}=2$ supergravity is to build up an $\mathcal{N}=2$ superconformal gauge theory.  We then turn this into a theory of supergravity by realizing the superconformal symmetries as spacetime symmetries (instead of internal symmetries).  The Weyl multiplet is the multiplet that contains all of the gauge fields of these superconformal transformations, as well as some auxiliary fields that must be added for consistency.  The bosonic content of the Weyl multiplet includes the metric $g_{\mu\nu}$, the dilatation generator $b_\mu$, the $SU(2)_R$ gauge field $\mathcal{V}_{\mu~j}^{~i}$ (where $i$ and $j$ are $SU(2)$ indices), the $U(1)_R$ gauge field $A_\mu$, the auxiliary anti-self-dual antisymmetric tensor $T^-_{\mu\nu}$ and the auxiliary real scalar $D$.

We will couple this Weyl multiplet to $n_V+1$ off-shell vector multiplets, indexed by $I = 0,\ldots,n_V$.  The bosonic content of each vector multiplet is a complex scalar $X^I$, a $U(1)$ gauge field $W^I_\mu$, and an auxiliary $SU(2)$ triplet of real scalars $Y^I_{ij}$.

We summarize the bosonic field content of the Weyl and vector multiplets in table~\ref{table:multiplet_boson}, as well as the Weyl and chiral weights of each of the fields.

\bgroup
\def\arraystretch{1.5}
\begin{table}[H]
\centering
\begin{tabular}{|c|c|c|c|c|c|c||c|c|c|}\cline{2-10}
	\multicolumn{1}{c|}{} & \multicolumn{6}{c||}{Weyl Multiplet} & \multicolumn{3}{c|}{Vector Multiplet}\\ \hline
	 Field & $g_{\mu\nu}$ & $b_\mu$ & $A_\mu$ & $\mathcal{V}_{\mu~j}^{~i}$ & $T^-_{\mu\nu}$ & $D$ & $X^I$ & $W_\mu^I$ & $Y_{ij}^I$ \\ \hline
	 Weyl weight & $-2$ & $0$ & $0$ & $0$ & $1$ & $2$ & $1$ & $0$ & $2$ \\ \hline
	 Chiral weight & $0$ & $0$ & $0$ & $0$ & $-1$ & $0$ & $-1$ & $0$ & $0$ \\ \hline
\end{tabular}
\caption{Bosonic content of the Weyl and vector multiplets, with the corresponding Weyl and chiral weights.}
\label{table:multiplet_boson}
\end{table}
\egroup

We will also couple our theory to a chiral multiplet.  We will eventually identify the fields in the chiral multiplet with various contractions of fields from the Weyl multiplet, in order to introduce higher-derivative corrections to the theory.  For now, we will keep the chiral multiplet fully general.  The bosonic content of this multiplet includes the complex scalars $\hat{A}$ and $\hat{C}$, a complex $SU(2)$ triplet of scalars $\hat{B}_{ij}$, and an anti-self-dual tensor $\hat{F}^-_{\mu\nu}$.  The Weyl and chiral weights $w$ and $c$ of the lowest-component scalar $\hat{A}$ are arbitrary, but we can express the weights of the other fields in terms of these weights, as shown in table~\ref{table:multiplet_chiral}.

\bgroup
\def\arraystretch{1.5}
\begin{table}[H]
\centering
\begin{tabular}{|c|c|c|c|c|}\cline{2-5}
	\multicolumn{1}{c|}{} & \multicolumn{4}{c|}{Chiral Multiplet} \\ \hline
	 Field & $\hat{A}$ & $\hat{B}_{ij}$ & $\hat{F}^-_{\mu\nu}$ & $\hat{C}$ \\ \hline
	 Weyl weight & $w$ & $w+1$ & $w+1$ & $w+2$ \\ \hline
	 Chiral weight & $c$ & $c+1$ & $c+1$ & $c+2$ \\ \hline
\end{tabular}
\caption{Bosonic content of the chiral multiplet, with arbitrary Weyl and chiral weights $w$ and $c$ for the lowest-component scalar $\hat{A}$.}
\label{table:multiplet_chiral}
\end{table}
\egroup

%%%%%%%%%%%%%%%%%%%%%%%%%%%%%%%%%%%%%%%%%%%%%%%%%%%%%%%%%%%%%%%%%%

\subsection{Off-Shell Action}

%%%%%%%%%%%%%%%%%%%%%%%%%%%%%%%%%%%%%%%%%%%%%%%%%%%%%%%%%%%%%%%%%%

The interactions between the Weyl multiplet and the matter fields in the vector and chiral multiplets are conveniently summarized by introducing a prepotential $F\equiv F(X^I,\hat{A})$, a meromorphic function of the vector multiplet scalars $X^I$ and the chiral multiplet scalar $\hat{A}$.  Derivatives of the prepotential are denoted by
\begin{equation}
	\frac{\partial F}{\partial X^I} = F_I~,\quad \frac{\partial F}{\partial \hat{A}} = F_A~.
\end{equation}
The prepotential is holomorphic and does not depend on the complex conjugate scalars $\bar{X}^I$ and $\bar{\hat{A}}$, and so $F_{\bar{I}} = F_{\bar{A}} = 0$.  The prepotential is also homogeneous of second degree with respect to Weyl-weighted scalings of $X^I$ and $\hat{A}$, so
\begin{equation}
	F(\lambda X^I,\lambda^w \hat{A}) = \lambda^2 F(X^I,\hat{A})~,
\end{equation}
where $w$ is the Weyl weight of the chiral multiplet scalar $\hat{A}$ and $\lambda$ is some arbitrary scaling constant. An equivalent statement of this homogeneity is
\begin{equation}
	F_I X^I + w F_A \hat{A} = 2F~.
\label{eq:prepot2}
\end{equation}

The action is
\begin{equation}
	\mathcal{S} = \int d^4x\,\sqrt{-g}\,\mathcal{L}~,
\end{equation}
where $\mathcal{L}$ is the Lagrangian for our off-shell theory.  The purely bosonic part of the Lagrangian that couples the Weyl multiplet, the vector multiplets, and the chiral multiplet via interactions dictated by the prepotential is
\begin{equation}\begin{aligned}
	8\pi\mathcal{L} &= \bigg{[} i D^\mu F_I D_\mu \bar{X}^I - i F_I \bar{X}^I\left(\frac{1}{6}R - D \right) -\frac{i}{8}F_{IJ}Y_{ij}^I Y^{Jij} \\
	&\quad + \frac{i}{4}F_{IJ}\left(F_{\mu\nu}^{-I} - \frac{1}{4}\bar{X}^I T_{\mu\nu}^-\right)\left(F^{-\mu\nu J} - \frac{1}{4}\bar{X}^J T^{-\mu\nu}\right) \\
	&\quad - \frac{i}{8} F_I \left(F_{\mu\nu}^{+I} - \frac{1}{4}X^I T_{\mu\nu}^+\right)T^{+\mu\nu} - \frac{i}{32}F T_{\mu\nu}^+ T^{+\mu\nu} \\
	&\quad + \frac{i}{2}F_{AI}\left(F_{\mu\nu}^{-I} - \frac{1}{4}\bar{X}^I T_{\mu\nu}^-\right)\hat{F}^{-\mu\nu}  + \frac{i}{2} F_A \hat{C} \\
	&\quad - \frac{1}{8}F_{AA}\left(\varepsilon^{ik}\varepsilon^{jl}\hat{B}_{ij}\hat{B}_{kl} - 2 \hat{F}_{\mu\nu}^- \hat{F}^{-\mu\nu}\right) - \frac{i}{4} F_{AI}\hat{B}_{ij}Y^{Iij}\bigg{]} +\text{h.c.}~,
\label{eq:lg}
\end{aligned}\end{equation}
where the (bosonic) supercovariant derivative acts on the vector multiplet scalars $X^I$ and the chiral multiplet scalar $\hat{A}$ by
\begin{equation}
	D_\mu X^I = (\partial_\mu - b_\mu -i A_\mu)X^I~, \quad D_\mu \hat{A} = (\partial_\mu - w b_\mu + i c A_\mu)\hat{A}~.
\label{eq:covderiv}
\end{equation}

The Lagrangian (\ref{eq:lg}) has a term linear in the auxiliary $D$ field
\begin{equation}
	8\pi\mathcal{L} = i(F_I \bar{X}^I - \bar{F}_I X^I)\left(D-\frac{1}{6}R\right) + ...~,
\end{equation}
which leads to inconsistent equations of motion.  In order to fix this, we can couple a non-linear multiplet to the Lagrangian such that all linear terms in $D$ are cancelled.  The bosonic content of this non-linear multiplet includes two $SU(2)$ scalar fields $\Phi^i_{~\alpha}$, where $i$ is the $SU(2)$ index and $\alpha = 1,2$, a real vector field $V_\mu$, and a complex antisymmetric matrix $M_{ij}$ of scalars.  Ignoring all fermionic terms, the non-linear multiplet is subject to the constraint
\begin{equation}
	D^\mu V_\mu -\frac{1}{2}V^\mu V_\mu - \frac{1}{4}|M_{ij}|^2 + D^\mu \Phi^i_{~\alpha}D_\mu \Phi^\alpha_{~i} = D + \frac{1}{3}R~,
\label{eq:nlconstraint}
\end{equation}
where $D^\mu V_\mu$ indicates the bosonic supercovariant derivative acting on $V_\mu$.  In order to cancel the linear $D$-dependence in (\ref{eq:lg}), we must add the term
\begin{equation}
	i(F_I \bar{X}^I - \bar{F}_I X^I)\left(D^\mu V_\mu -\frac{1}{2}V^\mu V_\mu - \frac{1}{4}|M_{ij}|^2 + D^\mu \Phi^i_{~\alpha}D_\mu \Phi^\alpha_{~i} - D - \frac{1}{3}R\right)
\end{equation}
to the Lagrangian.  The resulting bosonic Lagrangian is
\begin{equation}\begin{aligned}
	8\pi\mathcal{L} &= -\frac{i}{2} (F_I \bar{X}^I - \bar{F}_I X^I)R + \bigg{[} i D^\mu F_I D_\mu \bar{X}^I -\frac{i}{8}F_{IJ}Y_{ij}^I Y^{Jij} \\
	&\quad + \frac{i}{4}F_{IJ}\mathcal{F}^{-I}_{\mu\nu}\mathcal{F}^{-\mu\nu J} - \frac{i}{8} F_I \mathcal{F}^{+I}_{\mu\nu}T^{+\mu\nu} - \frac{i}{32}F T_{\mu\nu}^+ T^{+\mu\nu} + \frac{i}{2}F_{AI}\mathcal{F}^{-I}_{\mu\nu}\hat{F}^{-\mu\nu} \\
	&\quad + \frac{i}{2} F_A \hat{C} - \frac{1}{8}F_{AA}\left(\varepsilon^{ik}\varepsilon^{jl}\hat{B}_{ij}\hat{B}_{kl} - 2 \hat{F}_{\mu\nu}^- \hat{F}^{-\mu\nu}\right) - \frac{i}{4}F_{AI}\hat{B}_{ij}Y^{Iij}\bigg{]} +\text{h.c.} \\
	&\quad + i(F_I \bar{X}^I - \bar{F}_I X^I)\left(D^\mu V_\mu -\frac{1}{2}V^\mu V_\mu - \frac{1}{4}|M_{ij}|^2 + D^\mu \Phi^i_{~\alpha}D_\mu \Phi^\alpha_{~i} \right)~,
\label{eq:ltot}
\end{aligned}\end{equation}
where we have defined the supercovariant field strengths
\begin{equation}\begin{aligned}
	\mathcal{F}^{+I}_{\mu\nu} &= F^{+I}_{\mu\nu} - \frac{1}{4} X^I T^+_{\mu\nu}~, \\
	\mathcal{F}^{-I}_{\mu\nu} &= F^{-I}_{\mu\nu} - \frac{1}{4} \bar{X}^I T^-_{\mu\nu}~.
\end{aligned}\end{equation}

\subsection{Higher-Derivative Interactions}
At this point the bosonic Lagrangian (\ref{eq:ltot}) contains the Ricci scalar but no higher-derivative gravity terms.  One way to introduce these is to identify the chiral multiplet fields with various contractions of fields in the Weyl tensor, chosen precisely such that the supersymmetry variations are all consistent. Roughly speaking, we set the chiral multiplet to be the square of the Weyl multiplet. Ignoring all fermionic terms, this identifies the chiral multiplet fields as
\begin{equation}\begin{alignedat}{3}
	&\hat{A} & {} = {} & T^-_{\mu\nu}T^{-\mu\nu}~, \\
	&\hat{B}_{ij} & {} = {} & -16 \varepsilon_{k(i} \mathcal{V}_{\mu\nu~j)}^{~~k} T^{-\mu\nu}~, \\
	&\hat{F}^-_{\mu\nu} & {} = {} & {-16}\left(W_{\mu\nu\rho\sigma}T^{-\rho\sigma} + D T^-_{\mu\nu} + 2i A_{\rho[\mu}T^{-\rho}_{\nu]}\right)~, \\
	& \hat{C} & {} = {} & 32\left( W_{\mu\nu\rho\sigma}W^{\mu\nu\rho\sigma} + i{}^* W_{\mu\nu\rho\sigma}W^{\mu\nu\rho\sigma} + 6 D^2 - 2 A_{\mu\nu}A^{\mu\nu} - 2 A_{\mu\nu}\tilde{A}^{\mu\nu} \right. \\
	&&&\quad\quad - \frac{1}{2}T^{-\mu\nu} D_\mu D^\rho T^+_{\rho\nu} + \frac{1}{4}R^\mu_{~\nu}T^-_{\mu\rho}T^{+\nu\rho} + \frac{1}{256}T^-_{\mu\nu}T^{-\mu\nu}T^+_{\rho\sigma}T^{+\rho\sigma} \\
	&&&\left.\quad\quad + \frac{1}{2}\mathcal{V}_{\mu\nu~j}^{~~i}\mathcal{V}^{\mu\nu j}_{~~~~i} - \frac{1}{2}\mathcal{V}_{\mu\nu~j}^{~~i}\tilde{\mathcal{V}}^{\mu\nu j}_{~~~~i} \right)~,
\label{eq:chiralfields}
\end{alignedat}\end{equation}
where the field strength of the $SU(2)_R$ gauge field $\mathcal{V}_{\mu~j}^{~i}$ is
\begin{equation}
	\mathcal{V}_{\mu\nu~j}^{~~i} = \partial_\mu \mathcal{V}_{\nu~j}^{~i} - \partial_\nu \mathcal{V}_{\mu~j}^{~i} + \frac{1}{2}\mathcal{V}_{\mu~k}^{~i}\mathcal{V}_{\nu~j}^{~k} - \frac{1}{2}\mathcal{V}_{\nu~k}^{~i}\mathcal{V}_{\mu~j}^{~k}~.
\end{equation}
The scalar $\hat{A}$ now has Weyl weight $w = 2$ and chiral weight $c = -2$.  The bosonic Lagrangian (\ref{eq:ltot}) still retains the same form, but this identification introduces higher-derivative interactions to the theory that we are interested in studying.

%%%%%%%%%%%%%%%%%%%%%%%%%%%%%%%%%%%%%%%%%%%%%%%%%%%%%%%%%%%%%%%%%%

\section{Simplifying the Lagrangian}
\label{sec:simple}

%%%%%%%%%%%%%%%%%%%%%%%%%%%%%%%%%%%%%%%%%%%%%%%%%%%%%%%%%%%%%%%%%%

In this section, we will simplify the Lagrangian (\ref{eq:ltot}) both by partially gauge-fixing our theory and by eliminating various auxiliary fields via their equations of motion.

%%%%%%%%%%%%%%%%%%%%%%%%%%%%%%%%%%%%%%%%%%%%%%%%%%%%%%%%%%%%%%%%%%

\subsection{Partial Gauge-Fixing}
\label{subsec:partial_ind}

%%%%%%%%%%%%%%%%%%%%%%%%%%%%%%%%%%%%%%%%%%%%%%%%%%%%%%%%%%%%%%%%%%

The Lagrangian (\ref{eq:ltot}) has an $\mathcal{N}=2$ superconformal symmetry that acts as a gauge symmetry.  To obtain an $\mathcal{N}=2$ Poincar\'{e} supergravity theory, we must gauge-fix the extra gauge symmetries of the superconformal theory, including special conformal transformations, dilatations, and a local chiral $SU(2)_R\, \times \, U(1)_R$ symmetry.  We gauge-fix the special conformal symmetry by choosing the $K$-gauge
\begin{equation}
	b_\mu = 0~.
\label{eq:gauge_k}
\end{equation}
To gauge-fix the dilatational symmetry, we choose the $D$-gauge that sets the K\"{a}hler potential to be constant
\begin{equation}
	e^{-\mathcal{K}} \equiv i(F_I \bar{X}^I - \bar{F}_I X^I) = \frac{1}{\kappa^2}~,
\label{eq:gauge_d}
\end{equation}
with the value of the constant chosen to reproduce the standard normalization of the Einstein-Hilbert term in the action.  The local chiral $SU(2)_R$ invariance can be gauge-fixed by imposing the $V$-gauge
\begin{equation}
	\Phi^i_{~\alpha} = \delta^i_{~\alpha}~.
\label{eq:gauge_v}
\end{equation}
Finally, to gauge-fix the local chiral $U(1)_R$ symmetry, we choose the $A$-gauge
\begin{equation}
	X^0 = \bar{X}^0~.
\label{eq:gauge_a}
\end{equation}
Note that the $D$-gauge (\ref{eq:gauge_d}) and $A$-gauge (\ref{eq:gauge_a}) remove two degrees degree of freedom from the vector multiplet scalars, and thus the Poincar\'{e} supergravity theory has only $n_V$ independent scalars.  

%%%%%%%%%%%%%%%%%%%%%%%%%%%%%%%%%%%%%%%%%%%%%%%%%%%%%%%%%%%%%%%%%%

\subsection{Eliminating Auxiliary Fields}
\label{subsec:elim_ind}

%%%%%%%%%%%%%%%%%%%%%%%%%%%%%%%%%%%%%%%%%%%%%%%%%%%%%%%%%%%%%%%%%%

The remaining independent auxiliary fields in our theory are the $SU(2)_R$ gauge field $\mathcal{V}_{\mu~j}^{~i}$, the $U(1)_R$ gauge field $A_\mu$, the vector multiplet $SU(2)$ triplets $Y_{ij}^I$, the non-linear multiplet fields $V_\mu$ and $M_{ij}$, and the anti-self-dual antisymmetric tensor $T_{\mu\nu}^-$.  We will use the auxiliary equations of motion to eliminate everything except $A_\mu$ and $T_{\mu\nu}^-$ from the action.

If we insert the chiral multiplet field expressions (\ref{eq:chiralfields}) into the Lagrangian (\ref{eq:ltot}), we find that the fields $Y_{ij}^I$ and $\mathcal{V}_{\mu~j}^{~i}$ and their derivatives appear at least quadratically with one another in the action.  It is therefore consistent to set them both to zero 
\begin{equation}
	Y_{ij}^I = 0~, \quad \mathcal{V}_{\mu~j}^{~i} = 0~,
\end{equation}
at the level of the action.

We now want to eliminate the non-linear multiplet fields $V_\mu$ and $M_{ij}$ from (\ref{eq:ltot}), subject to the constraint (\ref{eq:nlconstraint}).  These non-linear multiplet fields interact with the other matter fields only through the K\"{a}hler potential $e^{-\mathcal{K}} = i(F_I \bar{X}^I - \bar{F}_I X^I)$, which is set to a constant via the $D$-gauge condition (\ref{eq:gauge_d}).  The non-linear multiplet fields effectively decouple from the rest of our theory, and so we can study their equations of motion independently from the others.  We find that we can choose
\begin{equation}
	V_\mu = 0~, \quad M_{ij} = 0~,
\end{equation}
at the level of the action, as long as the background value of $D$ satisfies $D = -\frac{1}{3}R$.

%%%%%%%%%%%%%%%%%%%%%%%%%%%%%%%%%%%%%%%%%%%%%%%%%%%%%%%%%%%%%%%%%%

\subsection{Resulting Lagrangian}
\label{subsec:resultingaction_ind}

%%%%%%%%%%%%%%%%%%%%%%%%%%%%%%%%%%%%%%%%%%%%%%%%%%%%%%%%%%%%%%%%%%

In subsections~\ref{subsec:partial_ind} and~\ref{subsec:elim_ind}, we found via partial gauge-fixing and elimination of auxiliary fields that we can consistently set
\begin{equation}
	b_\mu = Y_{ij}^I = \mathcal{V}_{\mu~j}^{~i} = V_\mu = M_{ij} = 0~,\quad \Phi^i_{~\alpha} = \delta^i_{~\alpha}~,
\label{eq:auxtrunc}
\end{equation}
at the level of the action.  This truncation requires that any solution to the theory satisfies the constraint
\begin{equation}
	\quad D = -\frac{1}{3}R~.
\end{equation}
Since the $SU(2)_R$ gauge field $\mathcal{V}_{\mu~j}^{~i}$ is set to zero, the chiral multiplet fields from (\ref{eq:chiralfields}) become
\begin{equation}\begin{alignedat}{3}
	&\hat{A} & {} = {} & T^-_{\mu\nu}T^{-\mu\nu}~, \\
	&\hat{B}_{ij} & {} = {} & 0~, \\
	&\hat{F}^-_{\mu\nu} & {} = {} & {-16}\left(W_{\mu\nu\rho\sigma}T^{-\rho\sigma} + D T^-_{\mu\nu} + 2i A_{\rho[\mu}T^{-\rho}_{\nu]}\right)~, \\
	& \hat{C} & {} = {} & 32\left( W_{\mu\nu\rho\sigma}W^{\mu\nu\rho\sigma} + i{}^* W_{\mu\nu\rho\sigma}W^{\mu\nu\rho\sigma} + 6 D^2 - 2 A_{\mu\nu}A^{\mu\nu} - 2 A_{\mu\nu}\tilde{A}^{\mu\nu} \right. \\
	&&&\left.\quad\quad - \frac{1}{2}T^{-\mu\nu} D_\mu D^\rho T^+_{\rho\nu} + \frac{1}{4}R^\mu_{~\nu}T^-_{\mu\rho}T^{+\nu\rho} + \frac{1}{256}T^-_{\mu\nu}T^{-\mu\nu}T^+_{\rho\sigma}T^{+\rho\sigma} \right)~.
\label{eq:chiralfields_2}
\end{alignedat}\end{equation}
If we take the full Lagrangian (\ref{eq:ltot}) and implement the consistent truncation (\ref{eq:auxtrunc}), we find
\begin{equation}\begin{aligned}
	8\pi\mathcal{L} &= -\frac{i}{2} (F_I \bar{X}^I - \bar{F}_I X^I)R + i D^\mu F_I D_\mu \bar{X}^I + \text{h.c.} \\
	&\quad + \bigg{[}\frac{i}{4}F_{IJ}\mathcal{F}^{-I}_{\mu\nu}\mathcal{F}^{-\mu\nu J} - \frac{i}{8} F_I \mathcal{F}^{+I}_{\mu\nu} T^{+\mu\nu} - \frac{i}{32}F T_{\mu\nu}^+ T^{+\mu\nu} \\
	&\quad + \frac{i}{2}F_{AI}\mathcal{F}^{-I}_{\mu\nu}\hat{F}^{-\mu\nu} + \frac{i}{2} F_A \hat{C} + \frac{i}{4}F_{AA} \hat{F}_{\mu\nu}^- \hat{F}^{-\mu\nu} \bigg{]} +\text{h.c.}~.
\label{eq:lgnew2}
\end{aligned}\end{equation}
This is the Lagrangian presented in (\ref{eq:lsimple}).

\flushbottom
\bibliographystyle{JHEP}
\bibliography{n2sugra_bibliography}

\end{document}